# Comment on “Surface roughness effect on ultracold neutron interaction with a wall and implications for computer simulations” A. Steyerl et.al., arxiv: 0911.4115v1 [1].

A.Serebrov, A.Fomin, V.Varlamov

*Petersburg Nuclear Physics Institute RAS*

**Abstract**

In above mentioned article [1] assumptions of possible systematic errors of our experiment on measuring of the neutron lifetime [2] have been made. In the given comment we are obliged to explain, that assumptions made in [1] are erroneous. They are caused by an inattentive reading of our work [2] and by the improper understanding the methods used in our work [2]. Simulation of experiment [2] carried out in [1] seems to be imperfect and doesn’t describe the experimental results. Therefore it is impossible to do any conclusions from such simulation.

## 1. Introduction

First of all it is necessary to emphasize that the main advantage of our work [2] is the small factor of UCN losses at storage in the traps. The probability of UCN losses is only 1 % from the probability of neutron β-decay. The storage time in the trap differs from the neutron lifetime on 5-15 sec depending on the UCN energy and sizes of a trap (see Fig. 12 from work [2]). To search for an error about 7 s (difference from world average value) in an extrapolation to a neutron lifetime is, at least, naive. The extrapolation with precision only 10 % provides the declared precision of experiment (±0.8 sec).

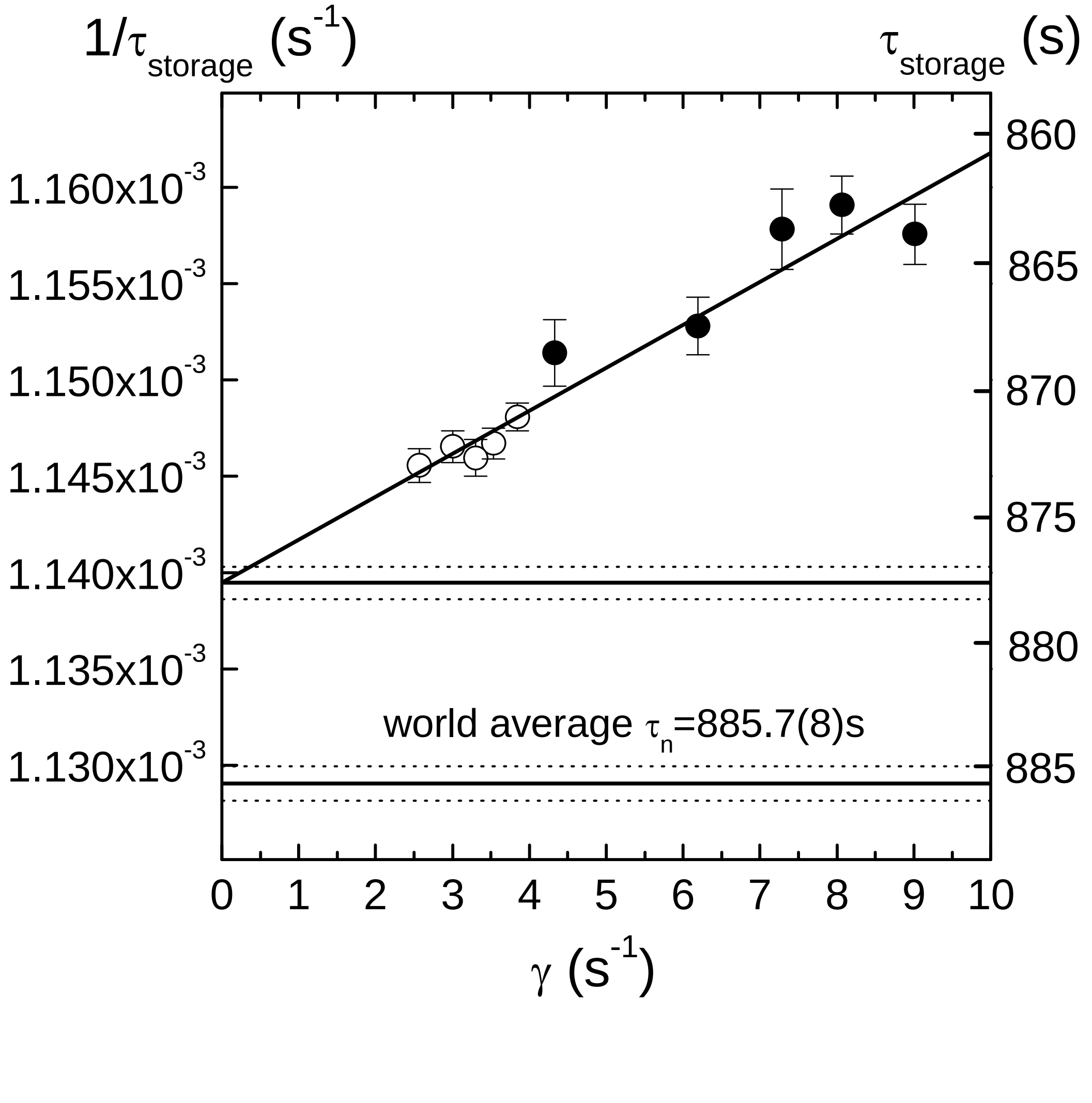

FIG. 12. Result of extrapolation to the neutron lifetime when combined energy and size extrapolations are used. The open circles represent the results of measurements for a quasispherical trap, and the full circles - the results of measurements for a cylindrical trap.

## 2. A method of size extrapolation.

In [1] the statement that “data interpretation for this experiment relied heavily on computer simulations” is made. Actually in our work a method of size extrapolation is used. This method is based on simple circumstance, that change of trap sizes changes the frequency of collisions with the trap walls. This method results in the radical suppression of influence of energy dependence of UCN losses on results of extrapolation of UCN storage time to the neutron lifetime. Below we bring quotation from our work [2] (see small font), devoted to this question.

”In reality, when gravity is present, complete exclusion of $\mu(E)$ is impossible because of the integral nature of Eq.(17). However, the residual effect of the dependence $\mu(E)$ on the neutron lifetime is negligible. It will be shown further that in our experiment on the neutron lifetime measurement, the contribution from uncertainty of this dependence does not exceed 0.144 s, whereas the statistical accuracy of the measurement was 0.7 s. To demonstrate this fact the UCN normalized loss rates $\gamma$ were calculated for a series of model dependencies $\mu(v)$:

(i) $\mu(v) \sim$ constant,

(iii) $\mu(v) \sim v/v_{lim}$,

(iii) $\mu(v) \sim (v/v_{lim})^2$, and

(iv) $\mu(v) \sim (v/v_{lim})^3$,

where $v_{lim}$ is the trap boundary velocity. A coefficient in these dependencies is not important because it is outside the integral in Eq.(17) and is canceled in Eq.(15). Then, the values of the neutron lifetime were obtained by the size extrapolation method for each corresponding model dependency $\mu_i(v)$. The differences between the neutron lifetime values for $\mu(v)$ from Eq.(10) and for the model dependencies $\mu_i(v)$ are as follows:

(i) -0.158 s,

(ii) -0:022 s,

(iii) +0:1 s, and

(iv) +0:217 s.

The linear dependence $v/v_{lim}$ is the most similar to the theoretical dependence $\mu(v)$, especially for small velocities. The mean square value of the numbers listed is 0:144 s and can be used as an estimation of the uncertainty of the neutron lifetime value owing to the uncertainty of the $\mu(v)$ function shape (see Table I). In spite of the fact that the model dependencies $\mu_i(v)$ are very different, their influence on the result of the extrapolation to the neutron lifetime is rather weak. Therefore, the size extrapolation method based on the idea of using two traps makes it possible to reduce substantially systematic errors that are caused by the uncertainty of our knowledge about the function $\mu(E)$.”

Thus, attempts to find the correction to the results of extrapolation to the neutron lifetime due to influence of a surface roughness on the loss factor are completely groundless.

## 3. About computer simulation.

In our work computer simulation is used for demonstration of validity of a method of data processing (see a part 4.F). For this purpose the experiment has been simulated with the preset value of the neutron lifetime. The same method of data processing as in the experiment has been applied to the results of simulation. The result of simulation is the count rate of the detector as a function of time. The experimental time diagrams are shown in Fig.2 and 3 in work [2]. The comparison of the simulated curves with experimental one is shown in Fig 14 in work [2].

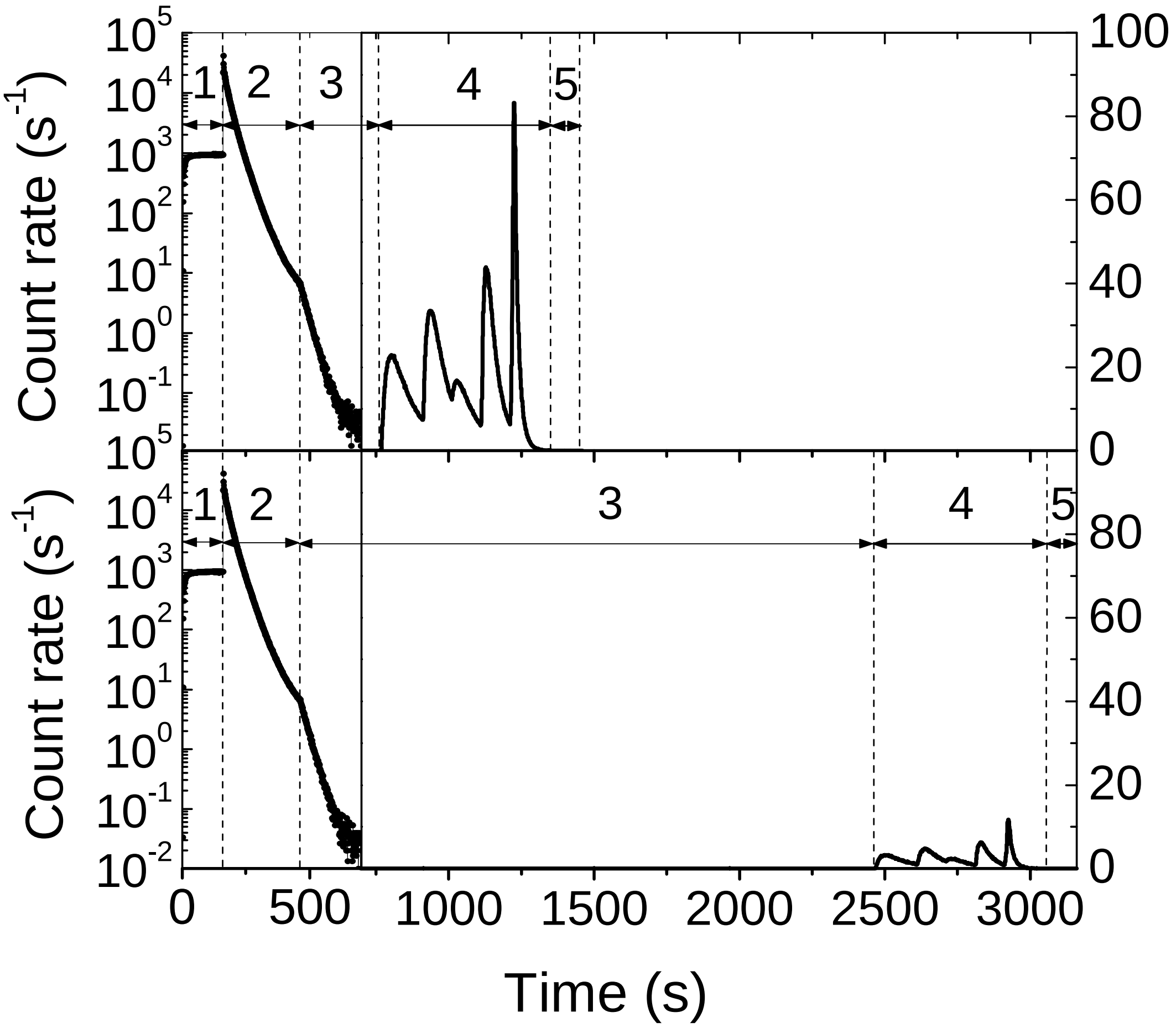

FIG. 2. Time diagrams of the storage cycle for two different holding times in a narrow trap: 1—filling, 160 s [with time of trap rotation (35 s) to monitoring position included]; 2—monitoring, 300 s; 3—holding, 300 s or 2000 s [with time of trap rotation (7 s) to holding position included]; 4—emptying, with five periods of 150, 100, 100, 100, and 150 s [with time of trap rotation (2.3, 2.3, 2.3, 3.5, and 24.5 s) to each position included]; and 5—measurement of background, 100 s.

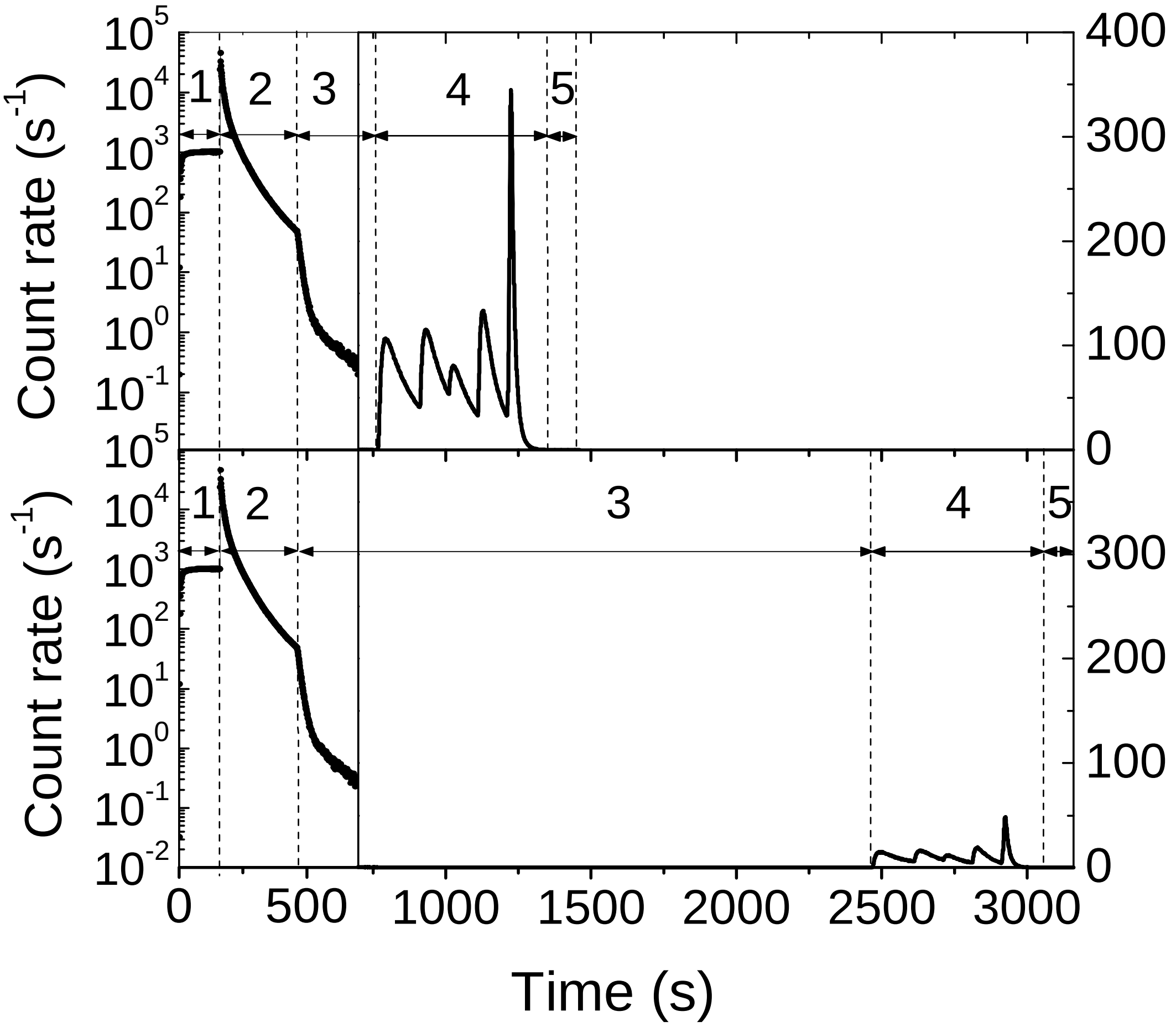

FIG. 3. Time diagrams of the storage cycle for two different holding times in a quasispherical trap: 1—filling, 160 s [with time of trap rotation (35 s) to monitoring position included]; 2—monitoring, 300 s; 3—holding, 300 or 2000 s [with time of trap rotation (7 s) to holding position included]; 4—emptying, with five periods of 150, 100, 100, 100, and 150 s [with time of trap rotation (2.3, 2.3, 2.3, 3.5, and 24.5 s) to each position included]; 5—measurement of background, 100 s.

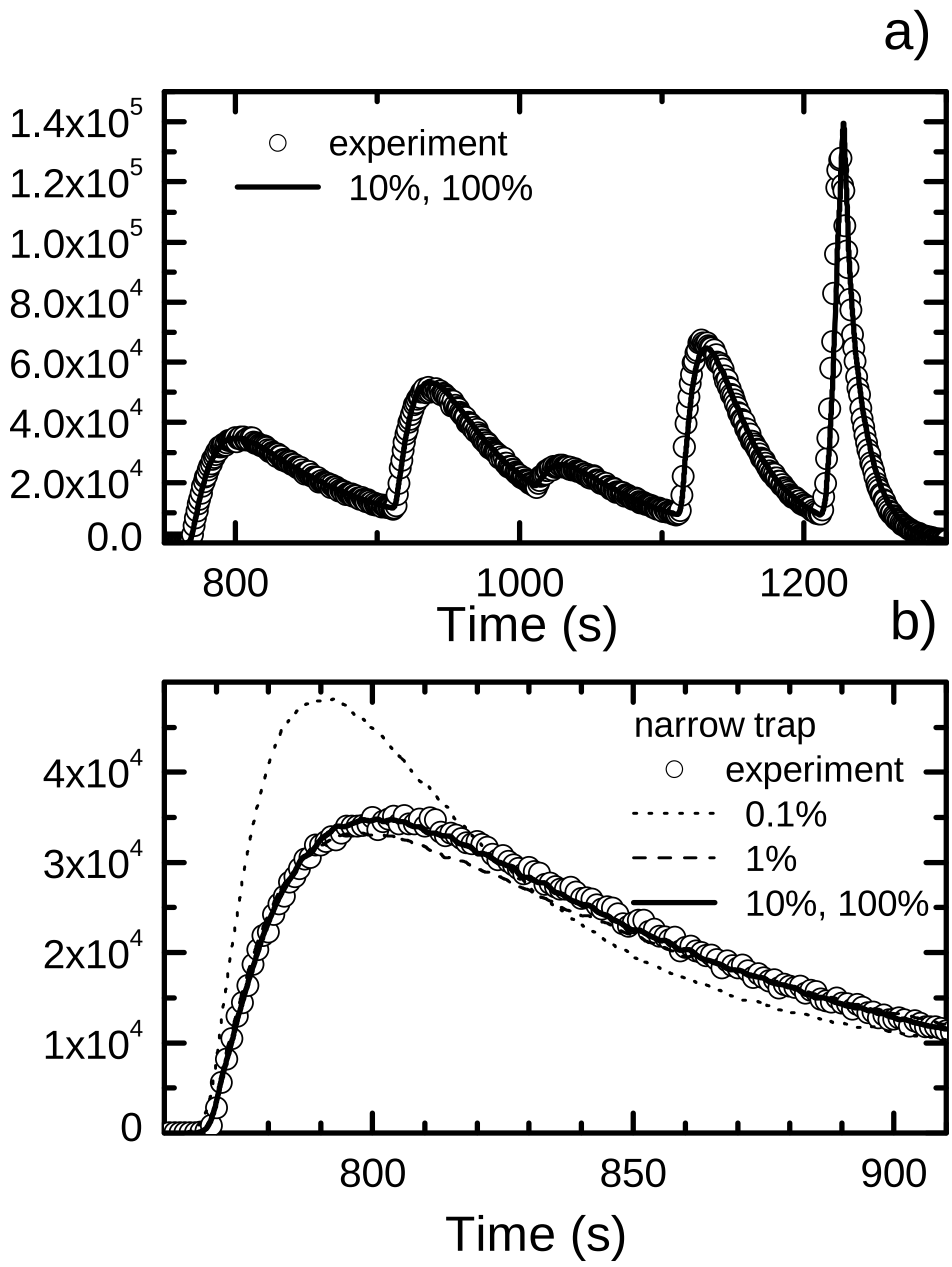

FIG. 14. Simulation of an experiment by the Monte Carlo method, consisting in simulating the neutron discharge from a narrow cylindrical trap. The dotted curve corresponds to the results of calculations with a 0.1% diffuse reflection probability; the dashed curve corresponds to a 1%diffuse reflection probability, and the solid curve to 10% and 100%.

As a result the preset value of the neutron lifetime has been restored with statistical accuracy of 0.236 sec. This accuracy has been accepted as the upper limit of possible systematic errors of the method of data processing. Results for neutron lifetime extrapolation of computer simulation of experiment are given on the Fig.15 from work [2]. They show the validity of both methods: the size extrapolation, as well as the energy extrapolation.

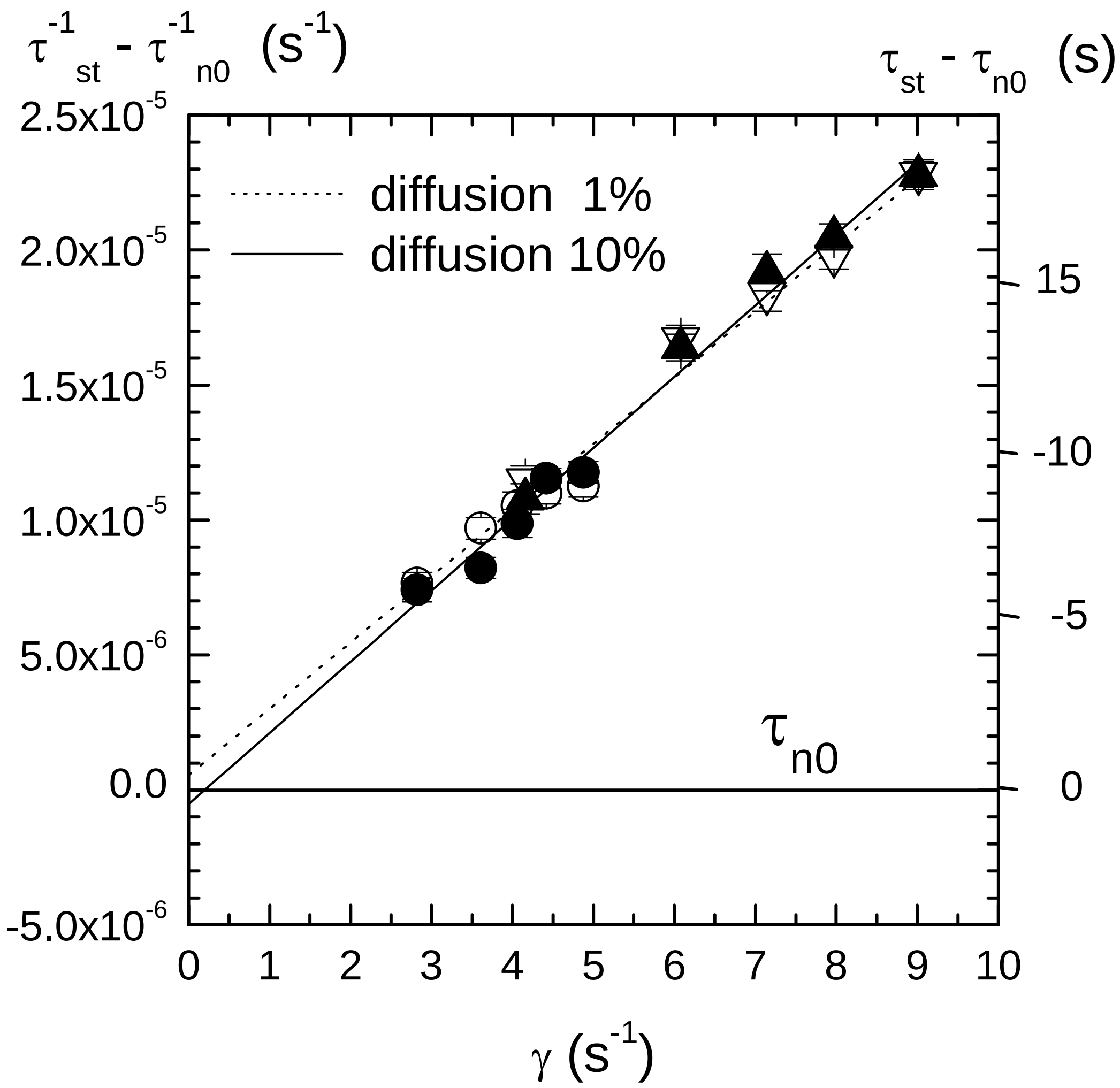

FIG. 15. Monte Carlo experiment with 1% and 10% diffuse reflection probability involving simulation of an extrapolation to the neutron lifetime. The circles represent the results of simulation for a wide trap and triangles the results of simulation for a narrow cylindrical trap. The open and full figures correspond to 1% and 10% diffuse reflection probability, respectively.

## 4. About diffuse scattering and the Lambert law

Work [1] without any grounds contains statement, that in our work [2] "isotropic" scattering was simulated. Below we bring a fragment of work [1] which contains this statement:

“A certain model of “isotropic” roughness scattering used in Monte Carlo simulations in [3,13] postulates that for each wall reflection there are two choices: Mirror reflection with a probability $P_m$ independent of incidence parameters k and $\theta_i$; or, with probability $(1-P_m)$, diffuse scattering uniformly distributed over solid angle.

Independence from k and $\theta_i$ contradicts (14), and since no weighting factor $\cos\theta$ is used this model violates detailed balance and is incompatible with an equilibrium isotropic distribution."

*(* In this place we would like to recommend for authors of work [1] to consult with colleagues before to publish in an open press the wrong accusations in our address and to mislead readers.)*

In order to demonstrate the importance of reflectivity model of diffuse scattering we have done a special test simulation before to simulate our experiment. The principle of comparison of the results of simulation is the same like presented in Fig.15. The results of this type of the test simulation are shown here in Fig. A(test).

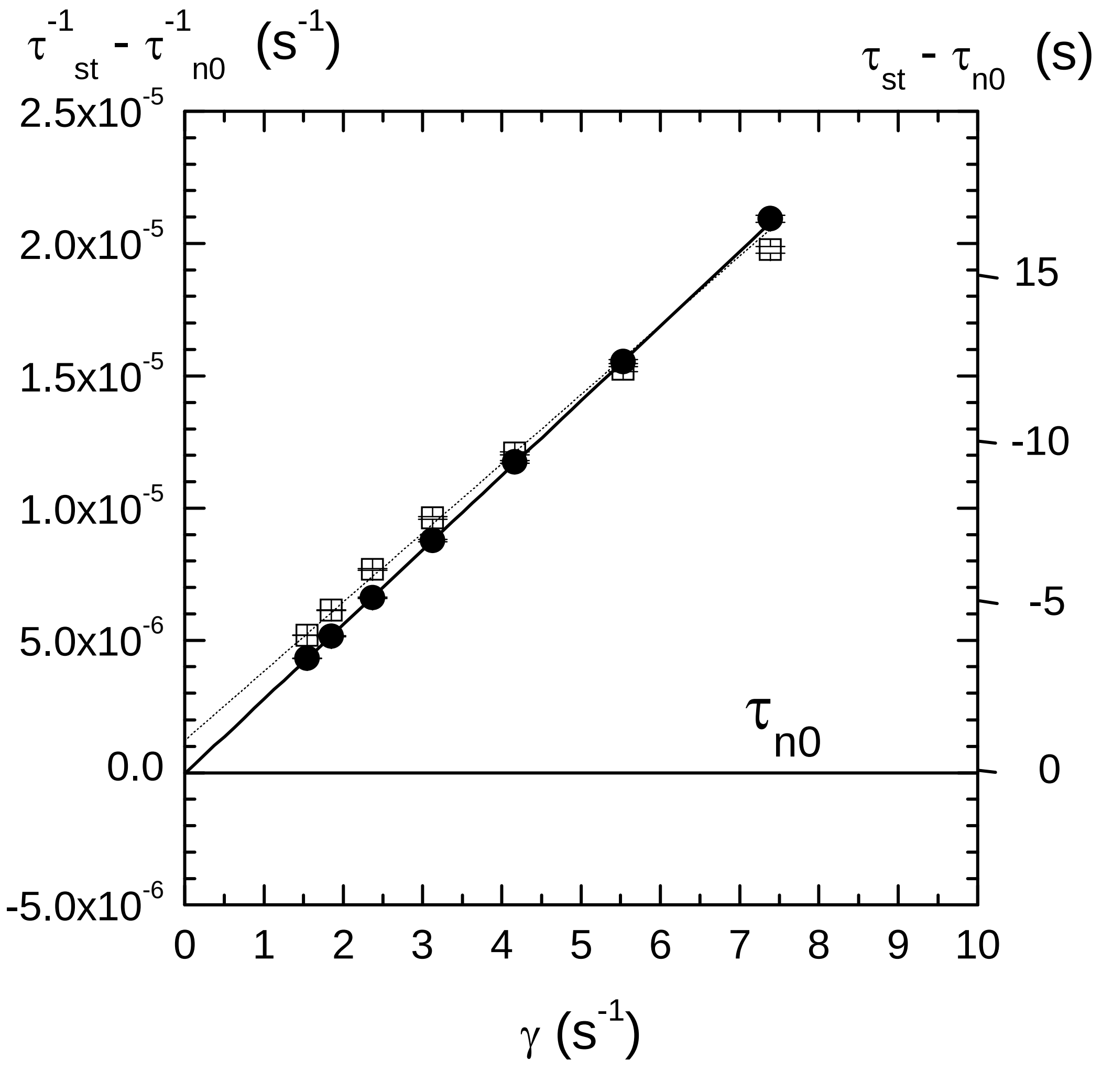

FIG. A(test). Monte Carlo experiment with different reflectivity models of diffuse scattering involving simulation of an extrapolation to the neutron lifetime. The filled circles represent the results of simulation when the model of diffuse scattering is described by the law of Lambert (correct simulation). The open squares the results of simulation when diffuse scattering uniformly distributed over solid angle (wrong simulation).

The test simulation was done for the cylindrical trap with the length of 13.5 cm and 76 cm in diameter. The boundary velocity of the trap surface is 6.8 m/s, the probability of the diffuse scattering is 100%. UCN losses were taken into account at each collision of UCN with the trap walls. Probability of UCN losses per bounce $\mu(v_{bis})$. depends from $v_{bis}$ - projection of UCN incident velocity vector to the bisector of angle between scattered and reverse incident velocity vectors. In case of specular reflection velocity $v_{bis}$ is simply projection of UCN incident velocity vector to the surface normal of substance. This model of the diffuse scattering is based on the simple assumption that reflection is specular one but surface in this point have the random orientation to satisfy Lambert law.

If the model of diffuse scattering is described by the law of Lambert (the results are shown with filled circles in Fig.A test) the preset value of the neutron lifetime has been restored with statistical accuracy of 0.015 sec and $\chi^2$ about unity. If we will use the model of diffuse scattering which uniformly distributed over solid angle (the results are shown with open squares in Fig. A test) then the preset value of the neutron lifetime has been restored with deviation about 1 sec and very poor $\chi^2$. Therefore the overall effect of the reflectivity model of diffuse scattering is not so big. Nevertheless using of the Lambert law is obligatory.

## 5. A question about existence of steady trajectories of neutron movement in a trap.

The special attention in our work [2] has been paid to a question about steady trajectories of movement of UCN in a trap at a high degree of smooth of a surface. Such effect could bring a problem to data processing of experiment. With this purpose the stability of a method has been checked up at a different probability of mirror reflection: 90 %, 99 % and 99.9 %. Other part of the reflected intensity (accordingly 10 %, 1 % and 0.1 %) was considered as a diffuse scattering.

Certainly, the diffuse scattering was described by the law of Lambert, i.e. the probability of a scattering is proportional to $\cos\theta$, where $\theta$ is a scattering angle in relation to a normal to surface of substance.

Results of our simulation with high probability of mirror reflection 99.9 % (accordingly diffuse 0.1 %) are presented on the Fig. 16 of work [2].

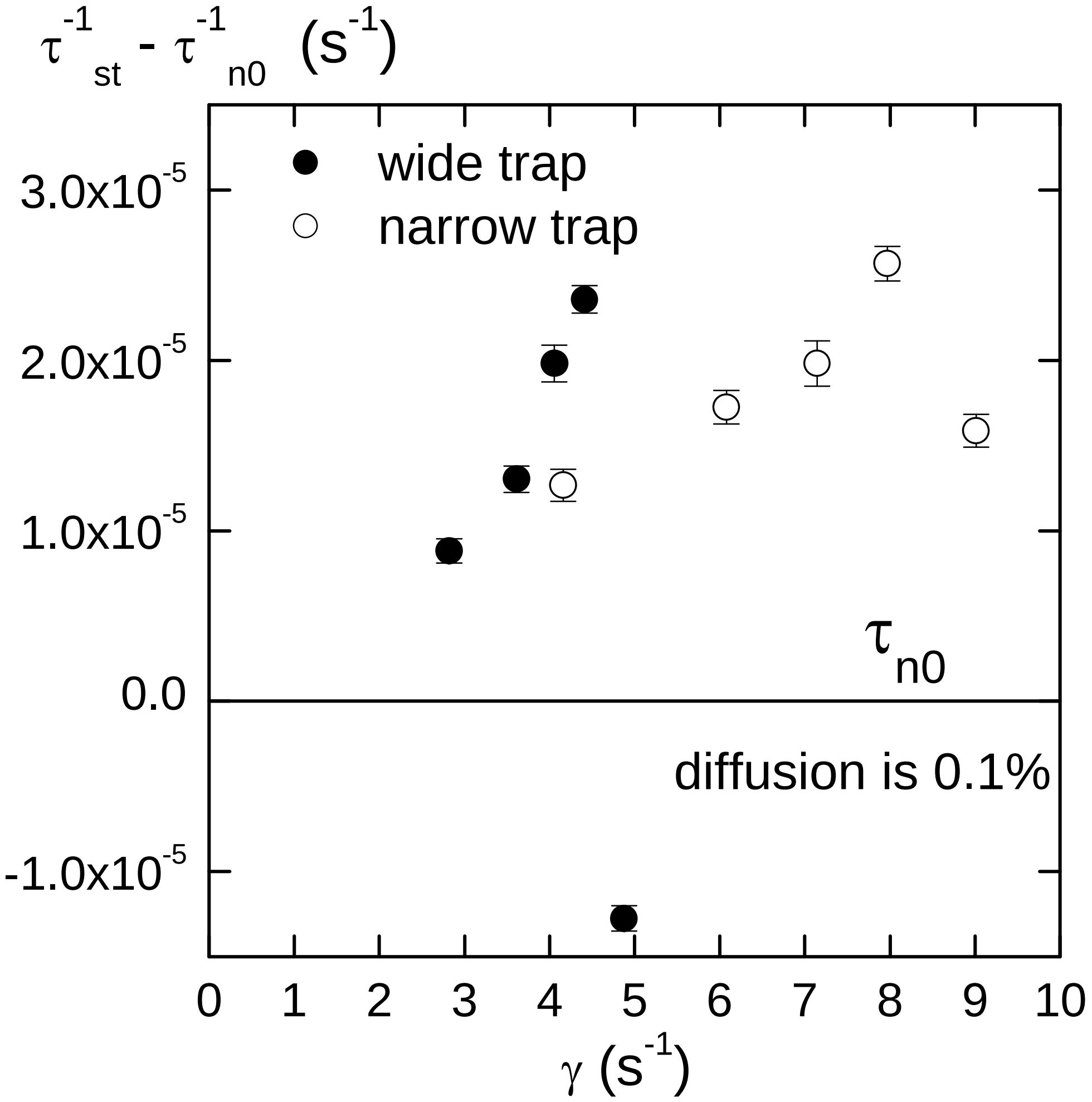

FIG. 16. Monte Carlo experiment with a 0.1% diffuse reflection probability involving simulation of an extrapolation to the neutron lifetime. The full circles represent the results of simulation for a wide trap and the open circles the results of simulation for a narrow cylindrical trap.

It is possible to see, that at so high probability of mirror reflection the stability of methods of extrapolation is broken. However, methods of extrapolation are steady at a fraction of a diffuse scattering of 1 % and 10 % that is visible on the Fig. 15 of work [2].

Basically, distribution of experimental data on the Fig. 12 in work [2] already shows that the case of high probability of mirror reflection is not realized in experiment.

It is necessary to notice, that absence of high probability of mirror reflection from the trap surface is obvious from the visual observation of an initial surface of a copper trap. Weld seams of the trap have been processed by an abrasive paper with the minimal size of a grain 10-15 μm. Accordingly the roughness of a surface can be observed visually even without a magnifying glass. When the trap surface was coated with PFPE at room temperature the impression of high mirror reflection of the surface does not arise, since the layer of PFPE basically repeats available roughness of a copper surface. However, visual observations do not give the quantitative performances. Therefore estimations of the probability of mirror reflection of a covering have been made directly in experiment with UCN.

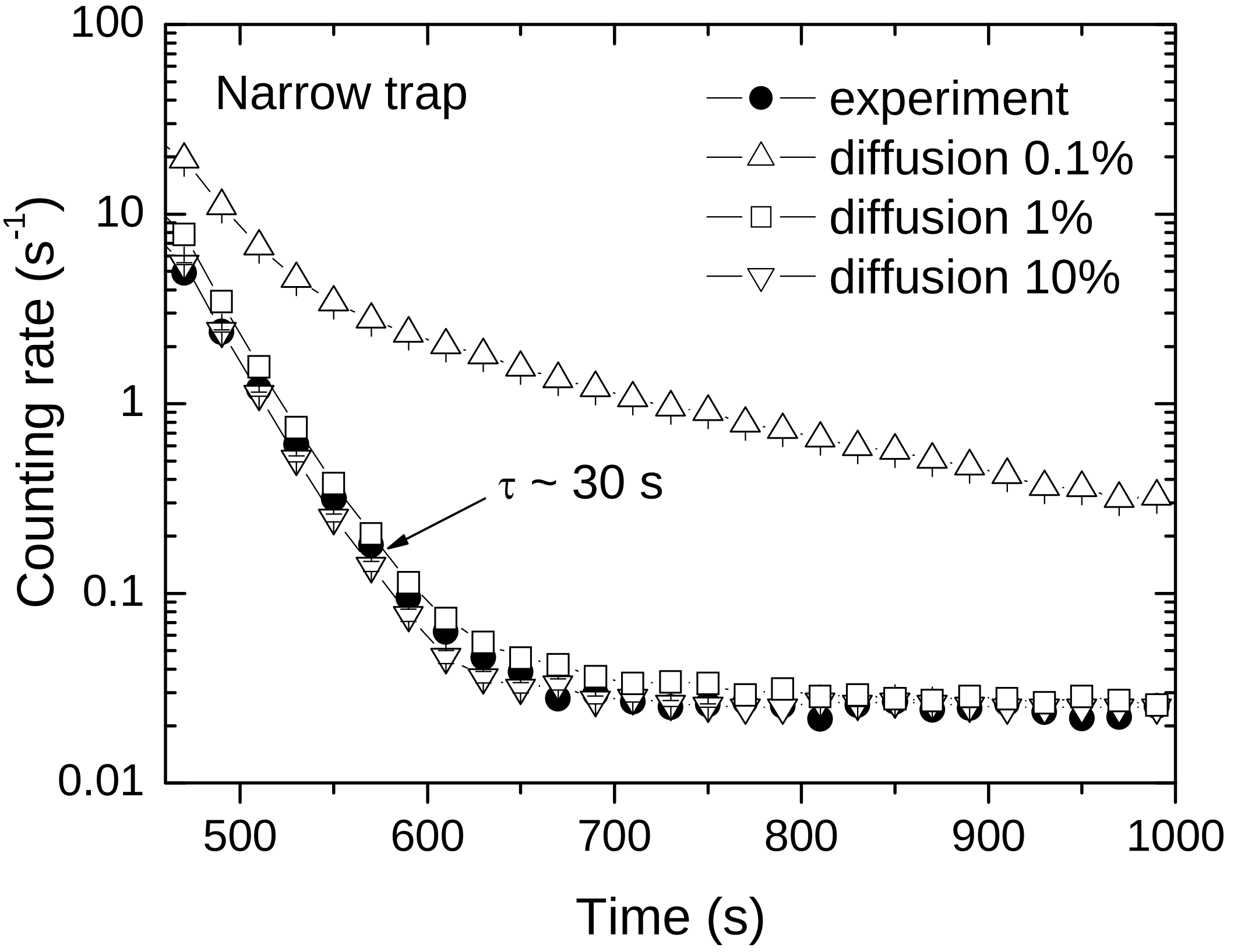

FIG. 17. Monte Carlo simulation of leakage process of UCN exceeding the gravitational barrier of the trap from the narrow cylindrical trap. The results of Monte Carlo calculations with different diffuse reflection probability are presented: 0.1% diffusion (open up triangles), 1% diffusion (open squares), and 10% diffusion (open down triangles). The filled circles represent the results of experiment.

The matter is that the quantity of UCN captured in a regime of steady mirror trajectories in a trap, and the speed of their flowing out depend on the probability of mirror reflection of the surface. With the help of computer simulation it has been shown, that process of flowing out of neutrons from the traps (observed in experiment after the beginning of a regime of storage) is successfully described at the probability of the diffuse scattering above 1 %. Results of simulation are presented on the Fig. 17 and 18 in work [2].

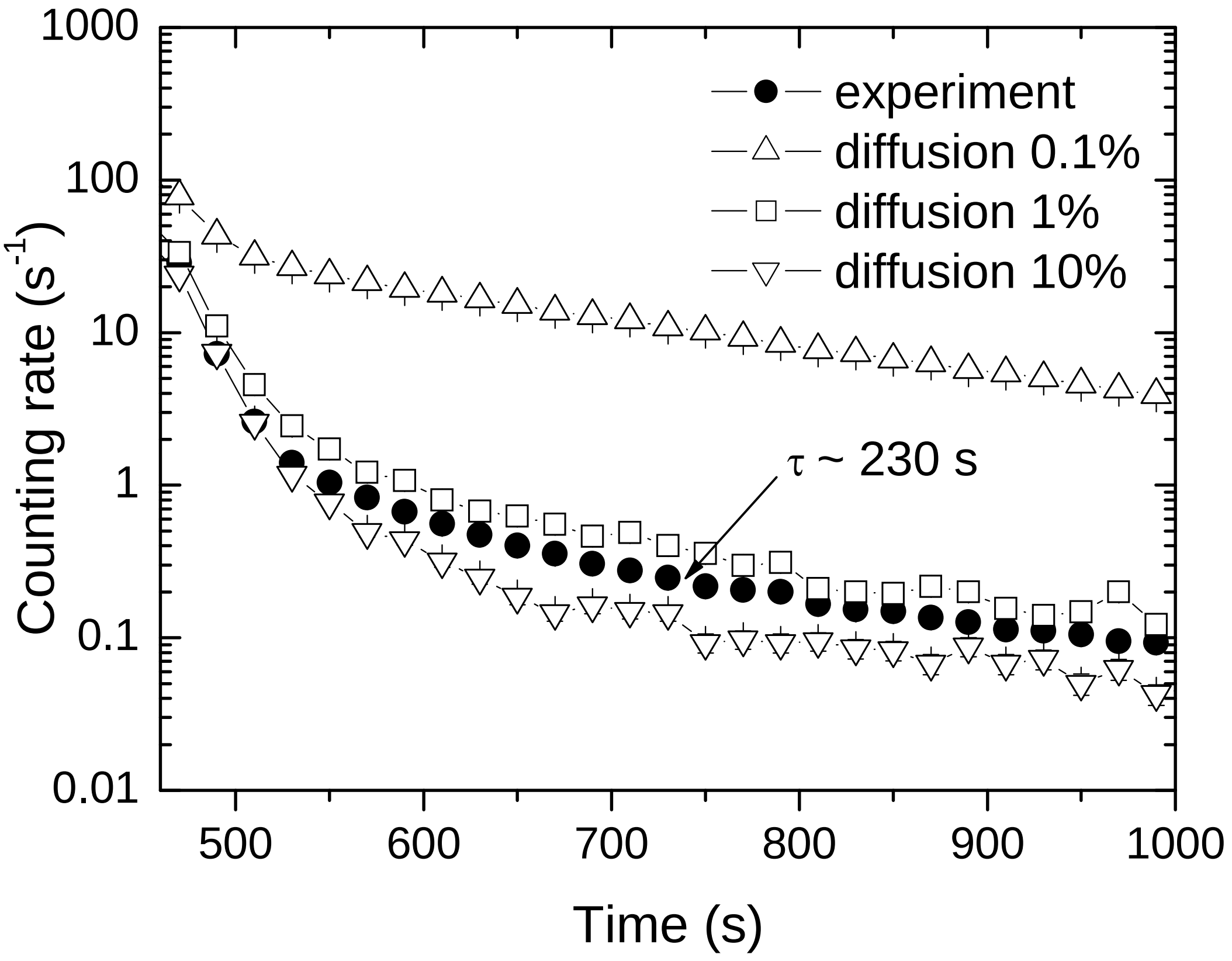

FIG. 18. Monte Carlo simulation of leakage process of UCN exceeding the gravitational barrier of the trap from the quasispherical trap. The results of Monte Carlo calculations with different diffuse reflection probability are presented: 0.1% diffusion (open up triangles), 1% diffusion (open squares), and 10% diffusion (open down triangles). The filled circles represent the results of experiment.

Thus it is reliably shown from the different sides, that the effect of steady mirror trajectories is absent in experiment [2]. Nevertheless, the question about the effect of steady mirror trajectories is put again in work [1], as though that the above-stated analysis in work [2] is not present. *(* we strongly recommend to authors of work [1] to study paragraph 4.F in work [2.])*

In work [1] independent simulation of the time diagram of our experiment is carried out. The results of this simulation are rather strange and logically not coordinated. (For example, the increase of diffusion scattering ($\alpha \approx 0.7$) increases the number of the high-energy UCN captured in the trap, whereas high-energy UCN should be captured more with increase of a smooth surface.) Simulation does not describe experimental dependences therefore any conclusions on the basis of such simulation are wrongful. Authors of work [1] should understand carefully all effects of own simulation before to do the unreasonable conclusions about possible systematic errors of experiment [2].